# A Frequency Domain Channel Estimation Based on Atomic Norm Minimization for Frequency Selective MmWave MIMO Systems

Mahdi Eskandari, Hamidreza Bakhshi

Department of electrical engineering, Shahed University, Tehran, Iran.

## ABSTRACT

In this study, a channel estimator for millimeter wave (mmWave) systems is proposed. By considering the sparse nature of channels in millimeter wave band, the channel estimation problem formulated as an atomic norm minimization problem. Previous studies on mmWave channel estimation have focused on frequency flat channels. However, in this study the work is based on a frequency selective channel and adopted atomic norm minimization and reweighed atomic norm minimization technique for channel estimation in the frequency domain. Simulation results verify the accuracy of the atomic and reweighted atomic norm minimization techniques.

*Keywords:*

channel estimation, millimeter wave, atomic norm minimization, reweighed atomic norm minimization, off grid channel estimation, MIMO.

## 1.INTRODUCTION

Millimeter wave communications is a promising technology for future wireless network due to its vacant spectrum resource [1]. Channel estimation in mmWave systems is of great importance, but most of existing schemes are focused on narrow-band flat fading channels while mmWave channels due to their considerably large bandwidth and different delays of multipath are frequency-selective and channel estimation in these systems is important. Because of poor scattering nature of signals in mmWave frequencies [2], MIMO mmWave channels are considered as a sparse problem and can be solved by on-grid Compressed Sensing (CS) tools [3], [4]. The main limitation of most of the prior studies is essentially from considering mmWave channels as a narrow-band frequency-flat channel since mmWave channels are wide-band frequency selective. There has been a number of studies considering mmWave channels as a frequency-selective channel which presented a few CS-based solutions to estimate them [5], [6], [7]. Recently, an off-grid CS approach was developed via Atomic Norm Minimization (ANM) [8], [9]. ANM has been used for mmWave channel estimation in some prior works such as [10], [11], [12]. As mentioned earlier, mmWave channels are actually frequency-selective but there is no prior work that focus on estimating frequency selective mmWave channels with the ANM point of view. To the best of our knowledge, this is the first paper to formulate the frequency-selective channel estimation problem for mmWave MIMO systems as an atomic norm minimization problem. Reweighted Atomic Norm Minimization (RAM) [13] has never been used to estimate frequency selective mmWave channels. Additionally, we compare our approach with on grid CS-based channel estimators. Simulation results verify the good performance of the ANM and RAM schemes. The paper follows the ensuing pattern. The presentation of system and channel model of a frequency-selective mmWave system is in section 2. Section 3 presents the formulations of mmWave frequency-selective channel estimator which can be solved using both ANM and RAM techniques. Simulation results are given in Section 4. Finally, in Section 5 the paper is draws a conclusion.

*notations*: $a$ is a scalar, $\mathbf{a}$ is a vector, $\mathbf{A}$ is a matrix, and $\mathbb{A}$ represents a set. For a vector or matrix, the transpose, complex conjugate and Hermitian are denoted by $(.)^T$, $(.)^*$ and $(.)^H$, respectively. $\mathbf{I}_N$ is the identity matrix of size $N \times N$. $\text{diag}(\mathbf{a})$ is a diagonal matrix with the diagonal entries constructed from $\mathbf{a}$. $\mathbf{A}^{-1}$ and $\mathrm{tr}(\mathbf{A})$ are respectively the inverse and the trace of $\mathbf{A}$. The operation $\mathrm{vec}(.)$ converts a matrix into a vector. $\mathbf{A} \otimes \mathbf{B}$ and $\mathbf{A} \circ \mathbf{B}$ are the Kronecker product and Khatri-Rao product of $\mathbf{A}$ and $\mathbf{B}$. $\mathbb{A} \geq 0$ means that $\mathbb{A}$ is positive semidefinite (PSD). A circularly symmetric complex Gaussian random vector with mean $\boldsymbol{\mu}$ and covariance $\mathbf{C}$ is denoted by $\mathrm{CN}(\boldsymbol{\mu}, \mathbf{C})$. $\mathbb{E}[.]$ denotes expectation.

## 2. SYSTEM AND CHANNEL MODEL

In this section, we present the mmWave MIMO-OFDM hybrid architecture-based system model, followed by a description of the adopted wideband mmWave channel model.

### 2.1 System Model

An OFDM based mmWave link is considered that utilizes $K$ subcarriers for communicating $N_s$ data stream between transmitter and receiver as shown in Figure 1. Similar to [6], zero padding (ZP) is used instead of cyclic prefixing to avoid distortion of training data during reconfiguration of RF circuitry. It is assumed that the number of antennas at the transmitter and receiver sides are $N_t$ and $N_r$ antenna elements, respectively. The number of RF chains in Both the transmitter and receiver side is $N_{RF}$. For training data transmission, a frequency-flat hybrid precoder and a combiner to reduce complexity are utilized. The training precoder for the $m$-th training frame is denoted by $\mathbf{F}_{T,m} \in \mathbb{C}^{N_t \times N_s}$. The discrete time complex base-band training signal at the $k$-th subcarrier from the $m$-th training frame can be written as

$$\mathbf{x}_m[k] = \mathbf{F}_{T,m}\mathbf{s}_m[k] \in \mathbb{C}^{N_t \times 1} \qquad (1)$$

where $\mathbf{s}_m[k]$ is $N_s \times 1$ transmitted symbol sequence at subcarrier $k$. Our presumption is that the transmitted symbols satisfy $\mathbb{E}[\mathbf{s}_m[k]\mathbf{s}_m^*[k]] = P/N_s\,\mathbf{I}_{N_s}$, where $P$ *is* total transmission power and $N_s = N_{RF}$.

### 2.2 Channel Model

It is presumed that the $N_r \times N_t$ channel is frequency selective with a delay tap of length $N_c$. Therefore the $d$-th delay tap of the channel can be expressed as [6], [7]

$$\mathbf{H}_d = \sum_{l=1}^{L} \alpha_l p(dT_s - \tau_l)\mathbf{a}_R(\phi_l)\mathbf{a}_T^H(\theta_l),$$
$$d = 0,...,N_c - 1, \qquad (2)$$

where $L$ is the number of paths, $p(\tau)$ is a raised cosine (RC) pulse for $T_s$ space signaling evaluated at $\tau$ seconds which includes the effect of shaping pulse and low-pass filters. $\alpha_l$ and $\tau_l$ are the gain and the delay of the $l$-th, respectively. $\phi_l$ and $\theta_l$ are the angles of arrival and departure (AoAs and AoDs) of the $l$-th path and finally $\mathbf{a}_R(\phi_l) \in \mathbb{C}^{N_r \times 1}$ and $\mathbf{a}_T(\theta_l) \in \mathbb{C}^{N_t \times 1}$ are the array response vectors for the received and transmitted antenna arrays. The channel model in (2) can be written in a compact form as

$$\mathbf{H}_d[k] = \mathbf{A}_R \boldsymbol{\Sigma} \mathbf{A}_T^H, \qquad (3)$$

where $\boldsymbol{\Sigma} = \mathrm{diag}(\alpha_1 p(dT_s - \tau_1),...,\alpha_L p(dT_s - \tau_L))$, and

$$\mathbf{A}_R = [\mathbf{a}_R(\phi_1),...,\mathbf{a}_R(\phi_L)] \in \mathbb{C}^{N_r \times L} \qquad (4)$$

$$\mathbf{A}_T = [\mathbf{a}_T(\theta_1),...,\mathbf{a}_T(\theta_L)] \in \mathbb{C}^{N_t \times L}. \qquad (5)$$

Using the model from (2), the frequency-domain channel for the $k$-th, the subcarrier can be expressed as:

$$\mathbf{H}[k] = \sum_{d=0}^{N_c-1} \mathbf{H}_d e^{-j\frac{2\pi kd}{K}}. \qquad (6)$$

By considering $\rho_{k,l} = \sum_{d=0}^{N_c-1} p(dT_s - \tau_l) e^{-j\frac{2\pi kd}{K}}$, (6) can be rewritten as:

$$\mathbf{H}[k] = \sum_{l=1}^{L} \alpha_l \rho_{k,l} \mathbf{a}_R(\phi_l)\mathbf{a}_T^H(\theta_l) = \mathbf{A}_R \boldsymbol{\Sigma}[k] \mathbf{A}_T^H \qquad (7)$$

where $\boldsymbol{\Sigma}[k] = \mathrm{diag}(\alpha_1 \rho_{k,1},...,\alpha_L \rho_{k,L}), k = 0,...,K-1$. Vectorizing the channel matrix (7) gives the following result

$$\mathrm{vec}(\mathbf{H}[k]) = (\mathbf{A}_T^* \circ \mathbf{A}_R)\boldsymbol{\gamma}_{k,l} \qquad (8)$$

with $\boldsymbol{\gamma}_{k,l} = [\alpha_1\rho_{k,1},...,\alpha_L\rho_{k,L}]^T, k = 0,...,K-1$. If the receiver applies a training combiner $\mathbf{W}_{T,m} \in \mathbb{C}^{N_r \times N_{RF}}$ for the $m$-th frame, the received signal at the $k$-th subcarrier of the $m$-th training frame is

$$\mathbf{y}_m[k] = \mathbf{W}_{T,m}^H \mathbf{H}[k]\mathbf{x}_m[k] + \mathbf{W}_{T,m}^H \mathbf{z}_m[k], \qquad (9)$$

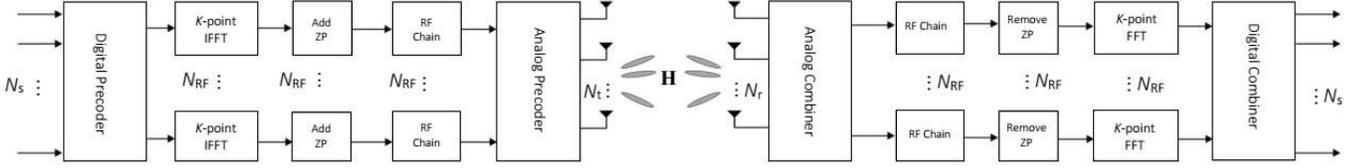

**Figure 1**: The structure of a mmWave MIMO-OFDM architecture.

where $\mathbf{z}_m[k] \sim \mathcal{CN}(\mathbf{0}, \sigma^2 \mathbf{I}_{N_r})$ is the circularly symmetric complex Gaussian distributed additive noise vector and SNR is defined as $\text{SNR} = \dfrac{P}{K\sigma^2}$.

## 3. CHANNEL ESTIMATION BASED ON ATOMIC NORM MINIMIZATION

Prior to formulating the problem in terms of atomic norm, firstly the concept of atomic norm needs to be elaborated. The mathematical theory of atomic norm was introduced in [14] and extended to line spectral estimation in [8]. The atomic $\ell_0$ norm of a signal $\mathbf{x}$ is defined as [14], [15]

$$\|\mathbf{x}\|_{\mathbb{A},0} = \inf_{\varphi_j, \theta_j, \beta_j} \left\{ J : \mathbf{x} = \sum_{j=1}^{J} \beta_j \mathbf{g}(\theta_j, \phi_j) \right\}, \quad (10)$$

where $\inf\{.\}$ is the infimum of the input set. The $\ell_0$ norm exploits sparsity to the highest possibility, but it is not convex and NP-hard to compute and cannot be globally solved with a practical algorithm. The convex relaxation of (10) is defined as a convex hull of $\mathbb{A}$ [14], [15]

$$\|\mathbf{x}\|_{\mathbb{A}} = \left\{ \varepsilon \geq 0 : \mathbf{x} \in \varepsilon \text{conv}(\mathbb{A}) \right\}$$
$$= \inf_{\phi_j, \theta_j, \beta_j} \left\{ \sum_{j=1}^{J} |\beta_j| : \mathbf{x} = \sum_{j=1}^{J} \beta_j \mathbf{g}(\theta_j, \varphi_j) \right\} \quad (11)$$

where $\text{conv}\left(\mathbb{A}\right)$ is the convex hull of $\mathbb{A}$. If we consider our class of signals as
$\mathbf{h}_v = (\mathbf{A}_T^* \circ \mathbf{A}_R) \boldsymbol{\gamma}_{k,l} = \sum_{l=1}^{L} \boldsymbol{\gamma}_{k,l} \mathbf{g}(\theta_l, \phi_l)$, the atom set will
be $\mathbf{g}(\theta_j, \phi_j) = \mathbf{a}_T^*(\theta_l) \otimes \mathbf{a}_R(\phi_l)$. It is proven in [15] and [16] that the equivalent 2D atomic norm can be formulated as follows

$$\|\mathbf{h}_v\|_{\mathbb{A}} = \inf_{U,\upsilon} \frac{1}{2N_r N_t} \text{tr}(\mathbb{S}(\mathbf{U})) + \frac{\upsilon}{2}$$
$$\text{s.t.} \quad \boldsymbol{\Xi} = \begin{bmatrix} \mathbb{S}(\mathbf{U}) & \mathbf{h}_v \\ \mathbf{h}_v^H & \upsilon \end{bmatrix} \geq 0, \quad (12)$$

where $\mathbf{U}$ is a matrix of size $\left(2N_t - 1\right) \times \left(2N_r - 1\right)$ and defined as

$$\mathbf{U} = [\mathbf{u}_{-N_r+1}, ..., \mathbf{u}_{N_r-1}], \quad (13)$$

where $\mathbf{u}_j = [\mathbf{u}(-N_t+1), ..., \mathbf{u}(N_t-1)]$ with $j = -N_r+1, ..., N_r-1$. $\mathbb{S}(\mathbf{U})$ is a Toeplitz matrix defined as

$$\mathbb{S}(\mathbf{U}) = \begin{bmatrix} \mathbf{T}(\mathbf{u}_0) & \cdots & \mathbf{T}(\mathbf{u}_{-N_r+1}) \\ \vdots & \ddots & \vdots \\ \mathbf{T}(\mathbf{u}_{N_r-1}) & \cdots & \mathbf{T}(\mathbf{u}_0) \end{bmatrix} \quad (14)$$

where $\mathbf{T}(.)$ denotes the Toeplitz matrix which its first column is the last $N_t$ elements of the input vector. In order to formulate the channel estimation problem as an atomic norm minimization problem, we should consider that the channel remains unchanged during $M$ training sequences, therefore the received signal can be written as

$$\mathbf{y}[k] = \boldsymbol{\Phi}[k] \mathbf{h}_v[k] + \mathbf{q}[k], \quad (15)$$

where
$\mathbf{y}[k] = [\mathbf{y}_1[k], ..., \mathbf{y}_M[k]]^T, \boldsymbol{\Phi}[k] = [\boldsymbol{\Phi}_1[k], ..., \boldsymbol{\Phi}_M[k]]^T$, and $\boldsymbol{\Phi}[k] = (\mathbf{x}_m^T[k] \otimes \mathbf{W}_m^H)$. According to [9] we can formulate the following optimization problem for estimating the channel

$$\hat{\mathbf{h}}_v[k] = \arg\min_{\mathbf{h}_v} \frac{1}{2} \left\| \mathbf{y} - \boldsymbol{\Phi}[k] \mathbf{h}_v[k] \right\|_2^2 + \zeta \left\| \mathbf{h}_v[k] \right\| \quad (16)$$

Using (12) and (16) the final atomic norm minimization problem can be written as a semi definite programming problem that can be solved using convex solvers.

$$\hat{\mathbf{h}}_{v,ANM}[k] = \arg \min_{\mathbf{h}_v, v, \mathbf{U}} \frac{1}{2} \left\| \mathbf{y} - \boldsymbol{\Phi}[k]\mathbf{h}_v[k] \right\|_2^2$$
$$+ \frac{\zeta}{2N_r N_t} \mathrm{tr}(\mathsf{S}(\mathbf{U})) + \frac{v\zeta}{2} \qquad (17)$$
$$\text{s.t.} \quad \boldsymbol{\Xi} = \begin{bmatrix} \mathsf{S}(\mathbf{U}) & \mathbf{h}_v \\ \mathbf{h}_v^H & v \end{bmatrix} \geq 0,$$

where $\xi$ is a regularization parameter and depends on noise power and defines as [9]

$$\zeta = \kappa(1 + \frac{1}{\log(N')}) \sqrt{N' \log(N') + N' \log(4\pi \log(N'))}, \qquad (18)$$

With $\kappa = \mathrm{E}[\mathbf{q}[k]\mathbf{q}^*[k]]$ and $N' = N_r N_t$. If the solution of (17) denoted by $\hat{\mathbf{h}}_{v,ANM}[k]$, then the estimated channel is $\hat{\mathbf{H}}[k] = \mathrm{vec}^{-1}(\hat{\mathbf{h}}_v[k])$. By repeating the same procedure for all of $K$ subcarriers, the frequency selective channel can be estimated. For enhancing sparsity, we can use reweighed atomic norm minimization (RAM) that was introduced in [13]. As we know there is an optimization problem in (10), however it promotes sparsity to the greatest extent possible, it is nonconvex and NP-hard to solve. but the relaxed version of it, i.e., (11) is convex and can be solved using convex solvers. With reweighed atomic norm minimization, the gap between two norms can be mitigated and the sparsity will be enhanced. The detailed implementation approach of RAM algorithm is given in [13] and the result is that the optimization problem in (17) can be reformulate as a RAM problem as:

$$\hat{\mathbf{h}}_{v,RAM}[k] = \arg \min_{\mathbf{h}_v, v, \mathbf{U}} \frac{1}{2} \left\| \mathbf{y} - \boldsymbol{\Phi}[k]\mathbf{h}_v[k] \right\|_2^2$$
$$+ \frac{\zeta}{2N_r N_t} \mathrm{tr}(\boldsymbol{\Theta}_j \mathsf{S}(\mathbf{U})) + \frac{v\zeta}{2} \qquad (19)$$
$$\text{s.t.} \quad \boldsymbol{\Xi} = \begin{bmatrix} \mathsf{S}(\mathbf{U}) & \mathbf{h}_v \\ \mathbf{h}_v^H & v \end{bmatrix} \geq 0,$$

where $\boldsymbol{\Theta}_j = (\mathsf{S}(\hat{\mathbf{U}}_{j-1}) + \xi \mathbf{I}_{N_r N_t})^{-1}$, $j = 1, ..., J$ and $\hat{\mathbf{U}}_0$ equals to $\hat{\mathbf{U}}$ that obtains from (17) and $J$ is the number of iterations. $\xi$ is a regularization parameter which causes the optimization problem in (19) play $\ell_0$ norm minimization as $\xi \to 0$ and $\ell_1$ norm minimization problem as $\xi \to 1$.

## 4. SIMULATION RESULTS

In this section, performance of ANM and RAM based frequency-selective channel estimation are investigated. In order to get the results, we used Monte Carlo simulation averaged over many independent realizations. The typical parameters of our model are as follows. We take $N_r = N_t = 16$ and $N_{RF} = 2$. A half-wavelength spacing uniform linear arrays (ULAs) used in both the transmitter and receiver side. So the steering vectors are

$$a_R(\varphi_t) = \frac{1}{\sqrt{N_r}} \left\{ e^{j\pi t \cos(\phi_t)} \right\}, t = 0, ..., N_r - 1 \quad \text{and}$$

$$a_R(\theta_t) = \frac{1}{\sqrt{N_r}} \left\{ e^{j\pi t' \cos(\theta_t)} \right\}, t' = 0, ..., N_r - 1. \text{ In this}$$

paper we consider fully connected phase shifting networks as described in [2]. We also consider the phase shifters used in both transmitter and receiver quantized with $Q = 7$ quantization bits. This means that the realizable angles in the phase shifters should be stack from the set $\mathbb{A}_{PS} = \left\{ \frac{2\pi Q'}{2^Q} \right\}, Q' = 0, ..., 2^Q - 1$. This implies that each entries of training precoder and combiner vector are of the form $[\mathbf{F}_{T,m}]_{i,j} = \frac{1}{\sqrt{N_t}} e^{j\omega_{i,j}}$ and

$[\mathbf{W}_{T,m}]_{i,j} = \frac{1}{\sqrt{N_r}} e^{j\omega'_{i,j}}$, $m = 1, ..., M$ with $\omega_{i,j}, \omega'_{i,j} \in \mathbb{A}_{PS}$. The number of OFDM subcarriers is $K = 32$. The number of channel paths is $L = 3$ and are independently and identically distributed with delay $\tau_l$ chosen at random from $\left[ 0, (N_c - 1)T_s \right]$, where $T_s$ is the sampling interval and set to be $\frac{1}{1760}$ $\mu sec$ as in the IEEE 802.11ad wireless standard. The roll-off factor of the raised cosine filter is $0.8$. The AoAs and AoDs are chosen uniformly at random from $[0, \pi)$. The number of delay taps of the channel is $N_c = 4$. Finally, we used CVX for solving (17) and (19). If the vectorized estimated channel denoted by $\hat{\mathbf{h}}$, our performance metric is Normalized Mean Squared Error (NMSE) and defined as [7]:

$$\mathrm{NMSE} = 10\log \left( \frac{\sum_{k=1}^{K} \left\| \hat{\mathbf{h}}[k] - \mathbf{h}[k] \right\|_2^2}{\sum_{k=1}^{K} \left\| \mathbf{h}[k] \right\|_2^2} \right) \qquad (20)$$

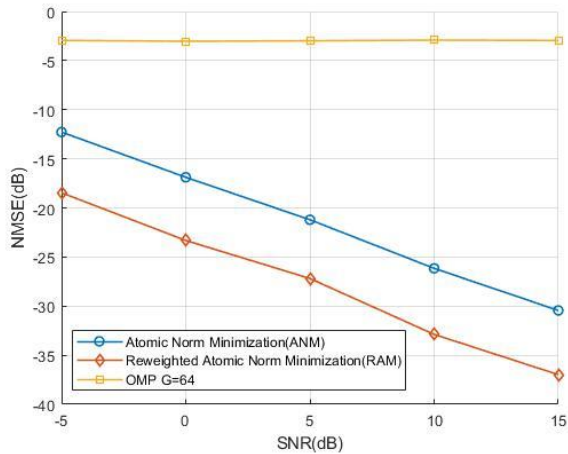

**Figure 2**: NMSE performance for the different algorithms with $M = 60$ number of training frames. For the case of on-grid OMP based channel estimator, both AoAs and AoDs quantized with $G$ points. The RAM algorithm implemented with $J = 5$ iterations and $\xi = 1$.

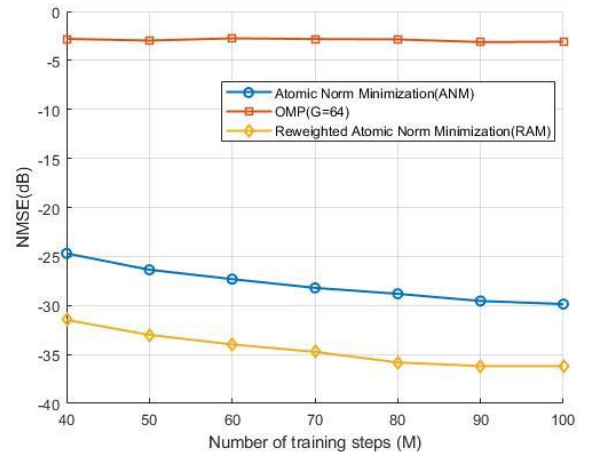

**Figure 3**: Evolution of NMSE vs. the number of training steps ( $M$ ) for the OMP, ANM and RAM algorithms with $\mathrm{SNR} = 10\mathrm{dB}$ .

The NMSE is our metric to compare different channel estimation schemes and will be averaged over many channel realizations. We compare our channel estimator with the on-grid OMP-based channel estimation technique that was proposed in [6].

Figure. 2 compares the average NMSE as a function of SNR for the ANM and RAM algorithms. It is obvious that the off-grid ANM and RAM algorithms estimate the frequency selective channels with more accuracy than the OMP-based estimator. Further, the RAM algorithm generates better performance than ANM because it promotes sparsity to the highest extent.

In Figure. 3, the performance of three channel estimators are compared as a function of the number of training steps. We assumed the same set of parameters as those used for generating Fig. 2. As we can see in the Fig. 3 with increasing training frames the RAM and ANM outperform the OMP based approach.

## 5. CONCLUSION

This study offers a channel estimator based on atomic norm minimization for mmWave MIMO-OFDM systems that exploits the sparsity of the channel in continuous angle of arrival and departure. For enhancing the sparsity, reweighed atomic norm minimization was used that promotes the sparsity to the greater extent. simulation results indicate that both channel estimators outperform the on-grid compressed sensing-based channel estimator.

## 1. Mahdi Eskandari

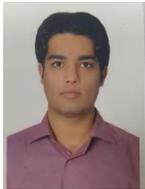

Mahdi Eskandari was born in Zanjan, Iran on January 21, 1994. He received his Bachelor degree in electrical engineering from university of Zanjan, Iran in 2016. Currently, he is a Msc. student in department of electrical engineering, Shahed University, Tehran, Iran. His research interests include signal processing, array signal processing, MIMO systems, and mmWave communication. E-mail: m-eskandari@shahed.ac.ir

## 2. Hamidreza Bakhshi

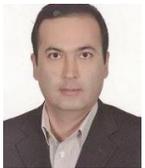

Hamidreza Bakhshi was born in Tehran, Iran on April 25, 1971. He received the B.Sc. degree in electrical engineering from Tehran University, Iran in 1992, and his M.Sc. and Ph.D. degree in Electrical Engineering from Tarbiat Modarres University, Iran in 1995 and 2001, respectively. |Since 2010, he has been an Associate Professor of Electrical Engineering at Shahed University, Tehran, Iran. His research interests include wireless communications, multiuser detection, and smart antennas. E-mail: bakhshi@shahed.ac.ir